\documentstyle[preprint,aps]{revtex}
\tighten
\begin{document}
\draft

\title{Some Thermodynamic Aspects of Black Holes and Singularities}

\author{C. O. LOUSTO\thanks{Electronic Address:
lousto@mail.physics.utah.edu}}
\address{Department of Physics, University of Utah,\\
201 JFB,
Salt Lake City, UT 84112, USA}
\date{\today}
\maketitle

\begin{abstract}
We review and correct the classical critical exponents characterizing
the transition from negative to positive black hole's heat capacity
at high charge--angular momentum. We discuss the stability properties
of black holes as a thermodynamic system in equilibrium with a
radiation bath (canonical ensamble) by using the Helmholtz
free energy potential. We finally analytically extend the analysis
to negative mass holes and study its thermodynamical stability behavior.

\end{abstract}

\section{Introduction}

The Kerr--Newman geometry (written in Boyer--Lindquist coordinates)
\begin{eqnarray}
ds^2=&-(\Delta-a^2\sin^2\theta)q^{-2}dt^2
-4Mra\sin^2\theta q^{-2}dtd\varphi+{q^2\over\Delta}dr^2\cr\cr
&+q^2d\theta^2+\left[(r^2+a^2)^2-\Delta a^2\sin^2\theta
\right]q^{-2}\sin^2\theta d\varphi^2~,
\label{kn}\end{eqnarray}
where
\begin{eqnarray}
\Delta=r^2-2Mr+a^2+Q^2\quad{\rm and}\quad q^2=r^2+a^2\cos^2\theta~,
\end{eqnarray}
represents the general stationary black Hole solution of Einstein
equations with an electromagnetic source. It is fully characterized by
three parameters: Total energy $M$, Charge $Q$, and Angular Momentum $J$.

The more important geometrical property of the black hole for us will be
its area $A_+$, defined as the area of the event horizon, located at a
radial coordinate $r_+$

\begin{equation}
A_+=4\pi(r_+^2+a^2)~,~~a={J\over M}~,
\label{area}\end{equation}
and where, more explicitly,
$r_\pm=M\pm\sqrt{M^2-a^2-Q^2}$ ($r_-$ being the internal or
Cauchy horizon).

The connection with thermodynamics can be made by the identification of
this geometrical quantity with the black hole entropy Ref. \cite{B73}
(we hereafter take units in which $G=1,$ $c=1$ and the Boltzmann constant,
$k_B=1$)

\begin{equation}
S_{BH}={1\over4}A_+~.
\label{entropy}\end{equation}

The independent derivation of the black hole radiation temperature
due to quantum effects in Ref. \cite{H74}, gave a firm foundation
to the thermodynamic
interpretation of the otherwise mechanical laws for black holes [It also
allowed to determine the exact proportionality factor in the above
equation.]

Inverting Eq (\ref{area}) we obtain the fundamental relation

\begin{equation}
M=\left({S\over 4\pi}+{\pi J^2\over S}+{Q^2\over 2}+{\pi Q^4\over4S}
\right)^{1/2}~,
\label{masa}\end{equation}

which upon differentiation allow us to obtain the {\it first law}
\begin{eqnarray}
dM=TdS+ \vec\Omega\cdot d\vec J +\Phi dQ~.
\label{first}\end{eqnarray}

Hence we identify the temperature, angular velocity and electric potential
\begin{equation}
T={\partial M\over\partial S}\biggr\vert_{JQ}
={1\over 4\pi}{r_+ - r_-\over(r_+^2+a^2)}~,~~
\Omega={\partial M\over\partial J}\biggr\vert_{SQ}
={a\over r_+^2+a^2}~,~~
\Phi={\partial M\over\partial Q}\biggr\vert_{SJ}
={Q r_+ \over  r_+^2 + a^2}~.
\label{temperatura}\end{equation}

Davies \cite{D77} was the first to observe that if we suppose that a
rotating charged black hole is held in equilibrium
at some temperature $T$, with a surrounding heat bath; the full
thermal capacity keeping $J$ and $Q$ constant,
\begin{equation}
C_{JQ}=T{\partial S\over\partial T}\biggr\vert_{JQ}={MTS^3\over\pi J^2+
{\pi\over4}Q^4-T^2S^3}~,\label{25}\end{equation}
goes from negative values for a Schwarzschild black hole,
$C_{Sch}=-M/T$, to positive values for a nearly extreme Kerr--Newman
black hole, $C_{EKN}\sim \sqrt{M^4-J^2-M^2Q^2} \to 0^+$
through an infinite discontinuity.

In Ref. \cite{L93} we tried to understand this divergence of the heat
capacity as indicative of critical phenomena in the thermodynamic
description of
black holes. We recall here that near the critical point (or curve in
our case) one can define critical exponents that characterize the behavior
of the relevant thermodynamic functions near criticality:

For the specific heat at constant $J$ and $Q$
\begin{equation}
C_{JQ}=T{\partial S\over\partial T}\biggr\vert_{JQ}\sim\cases
{(T_c-T)^{-\alpha'},  &~{\text{for}}~ $J=J_c$ ~{\text{and}}~ $Q=Q_c$ \cr
(J-J_c)^{-\varphi} ~{\text{or}}~ (Q-Q_c)^{-\varphi} &~{\text{for}}~
$T=T_c$\cr}~.
\label{19}\end{equation}

For the equation of state
\begin{equation}
\Omega-\Omega_c ~{\text{or}}~ \Phi-\Phi_c\sim\cases
{(T_c-T)^{\beta'},  & ~{\text{for}}~ $J=J_c$ ~{\text{and}}~ $Q=Q_c$ \cr
(J-J_c)^{1/\delta} ~{\text{or}}~ (Q-Q_c)^{1/\delta} & ~{\text{for}}~
$T=T_c$\cr}~.
\label{20}\end{equation}

For the isothermal capacitance or moment of inertia
\begin{eqnarray}
K^{-1}_{TQ,TJ}={\partial\Omega\over\partial J}\biggr\vert_{TQ}~{\text{or}}~
{\partial\Phi\over\partial Q}\biggr\vert_{TJ}\sim\cases
{(T_c-T)^{-\gamma'}, \quad\quad\quad ~{\text{for}}~
J=J_c ~{\text{and}}~& $Q=Q_c$ \cr
(J-J_c)^{1/\delta-1} ~{\text{or}}~ (Q-Q_c)^{1/\delta-1} ~{\text{for}}~&
$T=T_c$\cr}~.
\label{21}\end{eqnarray}

And for the entropy
\begin{equation}
S-S_c\sim\cases
{(T_c-T)^{1-\alpha'},  & ~{\text{for}}~ $J=J_c$  ~{\text{and}}~ $Q=Q_c$ \cr
(J-J_c)^{\psi} ~{\text{or}}~ (Q-Q_c)^{\psi} & ~{\text{for}}~ $T=T_c$\cr}~.
\label{22}\end{equation}

Primed exponents refer to temperatures below $T_c$, that is where heat
capacities are positive and black holes can be kept in equilibrium with
the radiation bath. Other two heat capacities, $C_{\Omega Q}$ and
$C_{J\Phi}$, also exhibit a singular behavior along two other critical
curves as shown in Ref. \cite{TL80,L93}.

The critical exponents defined above are not all independent but are
related by the scaling laws
\begin{eqnarray}
\alpha'+2\beta'+\gamma'=2~~,~~~\alpha'+\beta'(\delta +1)=2~,\nonumber
\end{eqnarray}
\begin{eqnarray}
\gamma'(\delta +1)=(2-\alpha')(\delta-1)~~,~~\gamma=\beta'(\delta -1)~,
\label{24}
\end{eqnarray}
\begin{eqnarray}
(2-\alpha')(\delta\psi-1)+1=(1-\alpha')\delta~,
{}~~\varphi+2\psi-\delta^{-1}=1~.\nonumber
\end{eqnarray}

These relations can be derived supposing the
thermodynamic potentials are homogeneous
functions of their variables near criticality. Since this law has
macroscopic important consequences and cannot be derived from the other
four, we call it the fourth law of thermodynamics.

The paper is organized as follows. In Sec. II we rederive the classical
critical exponents for black holes and rediscuss its equilibrium
properties in terms of the Helmholtz potential. In Sec. III we extend the
analysis to formally include negative mass objects. We end the paper
with a discussion of these results.

\section{Critical exponents and stability analysis}

{}From the first of Eqs. (\ref{temperatura}) and Eq. (\ref{masa})
we can obtain $T=T(S,J,Q)$, [For the sake of simplicity we will
collect the variables
into $x=(J,Q)$.] We then make the following Taylor expansion near the
critical curve
\begin{equation}
T(S,x)=T_c+{\partial T\over\partial S}\biggr\vert_{c,x}(S-S_c)
+{\partial T\over\partial x}\biggr\vert_{c,S}(x-x_c)+
{1\over2}{\partial^2T\over\partial S^2}\biggr\vert_{c,x}(S-S_c)^2+...
\label{serieT}\end{equation}

We first note at the black hole critical curve we have
${\partial T/\partial S}\big\vert_{c,x}=0$,
since that is $T/C_{JQ}\vert_c$
what vanishes at the critical point (see Ref. \cite{L93} for the explicit
expression of $T_c$)

Near the critical curve and for $x=x_c$, we then get
\begin{equation}
S-S_c\approx\pm\sqrt{2\over{-\partial^2T\over\partial S^2}
\big\vert_{c,x}}\sqrt{T_c-T}\quad\Rightarrow\quad\alpha'={1\over2},
\label{st}\end{equation}
where the last implication is obtained upon
comparison with Eq. (\ref{22}).

For the heat capacity we find
\begin{equation}
C_x\approx
T_c{\Delta S\over\Delta T}\biggr\vert_{x}\approx\text{Sign}[S_c-S]
\sqrt{2\over{-\partial^2T\over\partial S^2}\big\vert_{c,x}}
{T_c\over\sqrt{T_c-T}}\quad\Rightarrow\quad\alpha'={1\over2},
\label{ct}\end{equation}
this time the comparison is made with Eq. (\ref{19}).

Now letting be $T=T_c$ we get
\begin{equation}
S-S_c\approx\pm\sqrt{-2{\partial T\over\partial x}\big\vert_{c,S}
\over{-\partial^2T\over\partial S^2}\big\vert_{c,x}}
\sqrt{x_c-x}\quad\Rightarrow\quad\psi={1\over2},
\label{sx}\end{equation}
where now we looked at the definition of $\psi$ given in Eq. (\ref{22}).

And finally, for the heat capacity we obtain
\begin{equation}
C_x\approx T_c{\Delta S\over\Delta T}\biggr\vert_{x}\approx
\text{Sign}[S_c-S]
T_c\sqrt{2\over{\partial^2T\over\partial S^2}\big\vert_{c,x}\times
{\partial T\over\partial x}\big\vert_{c,S}}
{1\over\sqrt{x_c-x}}\quad\Rightarrow\quad\varphi={1\over2}.
\label{cx}\end{equation}
where we used the second row of Eq. (\ref{19}).

To find the other exponents we instead consider the Taylor development of
the equation of state. We shall use now the notation $y=(\Omega,\Phi)$ for
the conjugate variables of $x=(J,Q)$. Then we write
\begin{equation}
x(T,y)=x_c+{\partial x\over\partial y}\biggr\vert_{c,T}(y-y_c)+
{\partial x\over\partial T}\biggr\vert_{c,y}(T-T_c)
{1\over2}{\partial^2x\over\partial y^2}\biggr\vert_{c,T}(y-y_c)^2+...
\label{seriex}\end{equation}
Note at the black hole critical curve we have
${\partial x/\partial y}\big\vert_{c,T}=0$, since the inverse of
the isothermal capacitance or moment of inertia vanishes at the
critical point \cite{TL80}.

Again, near criticality, for $x=x_c$, we then get
\begin{equation}
y-y_c\approx\pm\sqrt{-2{\partial x\over\partial T}\big\vert_{c,y}
\over{-\partial^2x\over\partial y^2}\big\vert_{c,T}}
\sqrt{T_c-T}\quad\Rightarrow\quad\beta'={1\over2},
\label{yt}\end{equation}
where the last implication is found from Eq. (\ref{20}).

For the inverse of the moment of inertia or capacitance we find
\begin{equation}
K^{-1}_T\approx {\Delta y\over\Delta x}\biggr\vert_T\approx
\text{Sign}[y_c-y]
\sqrt{2\over{-\partial^2x\over\partial y^2}\big\vert_{c,T}\times
{\partial x\over\partial T}\big\vert_{c,y}}
{1\over\sqrt{T_c-T}}\quad\Rightarrow\quad\gamma'={1\over2},
\label{kt}\end{equation}
the comparison was made with Eq. (\ref{21}).

Now, taking $T=T_c$, we get
\begin{equation}
y-y_c\approx\pm\sqrt{2
\over{-\partial^2x\over\partial y^2}\big\vert_{c,T}}
\sqrt{x_c-x}\quad\Rightarrow\quad\delta=2,
\label{yx}\end{equation}
where we looked at Eq. (\ref{20}).

And finally, we obtain
\begin{equation}
K^{-1}_T\approx {\Delta y\over\Delta x}\biggr\vert_T\approx
\text{Sign}[y_c-y]
\sqrt{2\over{-\partial^2x\over\partial y^2}\big\vert_{c,T}}
{1\over\sqrt{x_c-x}}\quad\Rightarrow\quad\delta=2.
\label{kx}\end{equation}
where we made use of the second row of Eq. (\ref{21}).

The above considerations are general and apply to the charged and
rotating black holes (i.e. Kerr--Newman black holes represented by
metric (\ref{kn})). Although we have computed the exponents for the
canonical ensemble at $J,Q$ fixed the analysis
also apply to the other two
transitions we considered in Ref. \cite{L93}, at $\Omega,Q$ or
$J,\Phi$ fixed by properly identifying the variables $x$ and $y$
above and the thermodynamical potentials as in \cite{L93}.

We can check the correctness of the above derived
exponents by considering nonrotating charged black holes (i.e.
Reissner--Nordstr\"om ones) where equations simplify enough to
compute explicitly the behavior near the critical point. In particular,
it is easy to check form Eqs. (\ref{temperatura}) that the equation of
state can be written as
\begin{equation}
4\pi TQ=\Phi-\Phi^3~.
\label{ecest}\end{equation}

The critical values of the variables for the Reissner--Nordstr\"om
black hole are given by
\begin{equation}
\Phi_c={1\over\sqrt{3}}~,~~Q_cT_c={1\over6\pi\sqrt{3}}~.
\end{equation}
This allows us to write the equation of state as
\begin{equation}
(\Phi-\Phi_c)^2(\Phi+2\Phi_c)={1\over6\pi\sqrt{3}}
\left(1-{TQ\over T_cQ_c}\right)~,
\end{equation}
what near the critical point behaves like
\begin{equation}
\Phi-\Phi_c\approx\pm{1\over\sqrt{18\pi}}\cases
{(T_c-T)^{1/2},  &~{\text{for}}~ $Q=Q_c$ \cr
(Q_c-Q)^{1/2}, &~{\text{for}}~ $T=T_c$\cr}~.
\end{equation}
We thus see that $\beta'=1/2$ and $\delta=2$.

Using the above result for the behavior of $\Phi$ near criticality,
and given
\begin{equation}
S=\pi{Q^2\over\Phi^2}~,~~K^{-1}_{TQ}=-{4\pi T\over3\Phi^2-1}~,~~
C_Q={4\pi^2 Q^3\over\Phi^3(3\Phi^2-1)}~,
\label{entropia}\end{equation}
it is easy to check that all the exponents we get are those given by Eqs.
(\ref{st})--(\ref{kx}).

Let us now analyze the stability properties of black holes in equilibrium
with the thermal bath. The appropriate potential to study the canonical
ensamble is the Helmholtz free energy
\begin{equation}
F(T,\vec J,Q)=M-TS=TS+2\vec\Omega\cdot\vec J+\Phi Q~,
\end{equation}
where in the last equality we have used the mass formula of
Ref. \cite{S73}.

{}From this potential can be derived all the thermodynamical
variables of our interest
\begin{equation}
\Phi={\partial F\over\partial Q}\biggr\vert_{TJ}~,~~
\Omega={\partial F\over\partial J}\biggr\vert_{TQ}~,~~
S=-{\partial F\over\partial T}\biggr\vert_{JQ}~.
\label{primera}\end{equation}
and
\begin{equation}
C_{JQ}=-T{\partial^2F\over\partial T^2}\biggr\vert_{JQ}~,~~
K_{TQ}^{-1}={\partial^2 F\over\partial J^2}\biggr\vert_{TQ}~,~~
K_{TJ}^{-1}=J{\partial^2 F\over\partial Q^2}\biggr\vert_{TJ}~.
\end{equation}
{}From  Eqs. (\ref{st})--(\ref{kx}) we see that near criticality, first
derivatives of the potential are continuous (hence also the mass, what
implies zero latent heat), while second derivatives diverge.

Now, by use of the equation of state we can, in principle, rewrite
part of the Helmholtz potential as a function $f$ of the variables
$T,$ $\vec\Omega,$ and $\Phi$, in such a way that
\begin{equation}
F=f(T,\vec\Omega,\Phi)+\vec\Omega\cdot\vec J+\Phi Q~.
\end{equation}
We thus arrived to a form that looks like the effective potential
of an antiferromagnet (due to the plus sign of the last two addends)
in an external field $x=(Q,\vec J)$ and having an order parameter
$y=(\Phi,\vec\Omega)$ with two (or in general four) components,
which is the conjugate variable to $x$, i.e.
\begin{equation}
y={\partial F\over\partial x}\biggr\vert_{T}~,
\end{equation}
corresponding to the first two equations of (\ref{primera}).

The equation of state is given by the extremum condition of the
potential
\begin{equation}
{\partial F\over\partial y}\biggr\vert_{T}=0=
{\partial f\over\partial y}\biggr\vert_{T}+x~.
\end{equation}

Again, to be more explicit we can deal with the simpler case of
nonrotating charged black holes. In this case by use of
Eqs. (\ref{entropia}) and (\ref{ecest}) we obtain
\begin{equation}
F(T,\Phi,Q)={(1-\Phi^2)^2\over16\pi T}+Q\Phi~.
\label{potencial}\end{equation}

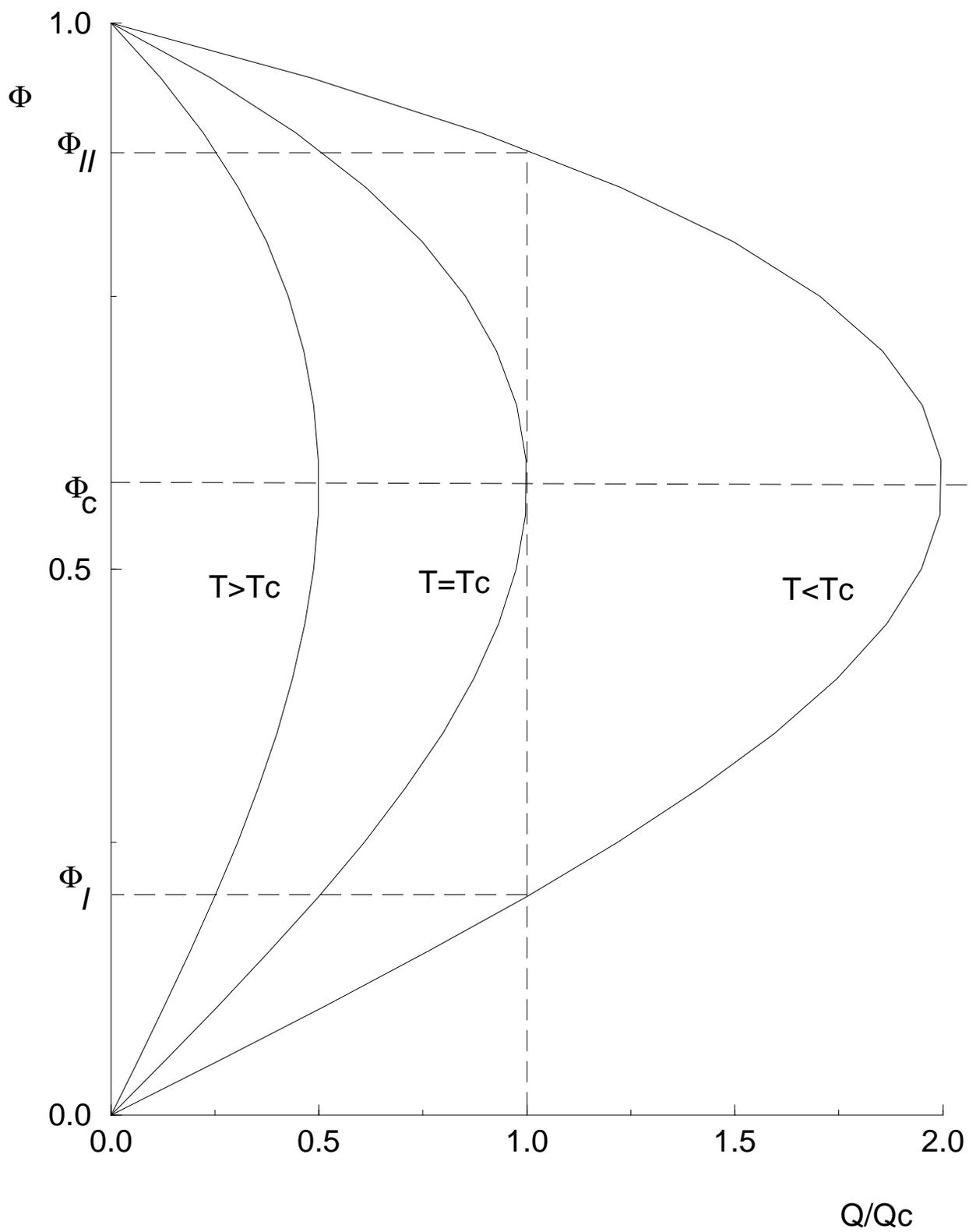
\begin{figure}
\caption{Isotherms for the Reissner--Nordstr\"om black hole. The
dashed line gives the location of the maximum always occurring at
$\Phi=1/\protect\sqrt{3}$.
For a given constant value of the charge $Q>0$ we can
have two, one or none values of $\Phi$. The low $\Phi$ branch corresponds
to black holes with $C_Q<0$; while the high $\Phi$ corresponds to those
with $C_Q>0$. The critical temperature is given by
$T_c=(6\pi\protect\sqrt{3}Q)^{-1}$.
For $Q<0$ we have the same figure reflected trough the origin to the third
quadrant.}
\label{fig1}
\end{figure}

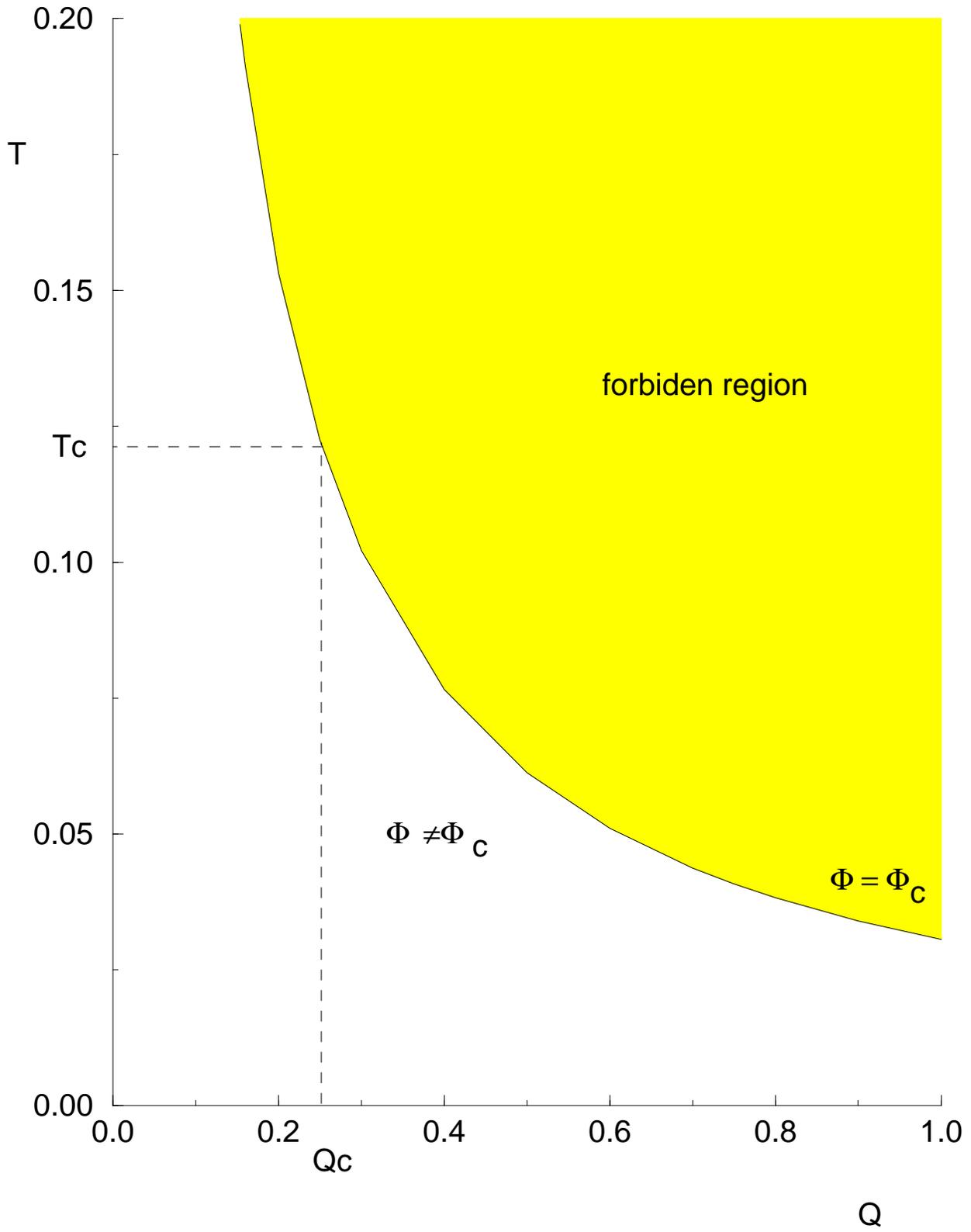
\begin{figure}
\caption{The equation of state of a nonrotating charged black hole
for different
slices of the electrostatic potential. For a fixed value of the charge,
the temperature is constrained to be less than the critical one. The
same is true for the charge if we fix the temperature.}
\label{fig2}
\end{figure}

We have plotted the equation of state (\ref{ecest}) in figures (\ref{fig1})
and (\ref{fig2}). In the first figure we fist observe that the charge
presents a maximum at the same value $\Phi_c=1/\sqrt{3}$,
its value being independent
of the temperature. If we consider a fixed value of the charge, like the
one shown in the figure, we see that at low temperatures there are two
possible values of $\Phi$, denoted by $I$ and $II$. As we rise the 
temperature of the system, those two states merge together at the critical
temperature $T_c=1/(6\pi\sqrt{3}Q)$. Above this temperature, no
equilibrium state is possible. One can also see that a state in the
branch $I$ of the equation of state has a higher mass than the
corresponding state labeled by $II$ since $\Phi_{II}>\Phi_I$ and from
Eqs. (\ref{temperatura})and (\ref{ecest}) we have
\begin{equation}
\Delta M=M_{II}-M_{I}={\Phi_I^4-\Phi_{II}^4\over T}\leq0~.
\end{equation}
For completeness,
the electrostatic potential can be found by applying Cardano's
formulae to the equation of state (\ref{ecest})
\begin{equation}
\Phi_I={2\over\sqrt{3}}\cos\left({\theta\over3}\right)~,~~
\Phi_{II}={2\over\sqrt{3}}\cos\left({\theta+4\pi\over3}\right)~,
\end{equation}
where $\cos{(\theta)}=-(6\pi\sqrt{3}TQ)$. $\Delta M$ ranges form
$-1/(8\pi T)$ for $Q=0$ to zero for $Q=Q_c$.

Finally, when the charge of the
black hole is negative we have the same plot reflected thought the origin
to the third quadrant.

In the second figure we see that
at a fixed value of the charge we can have any temperature from zero to
$T_c$ by letting $\Phi$ vary in the range (0,1) [The same happens for $T$
fixed; we have $0<Q<Q_c$.] i.e. $QT\leq1/(6\pi\sqrt{3})=Q_cT_c$.
When negative charges are considered we have
reflection of this figure to the fourth quadrant.

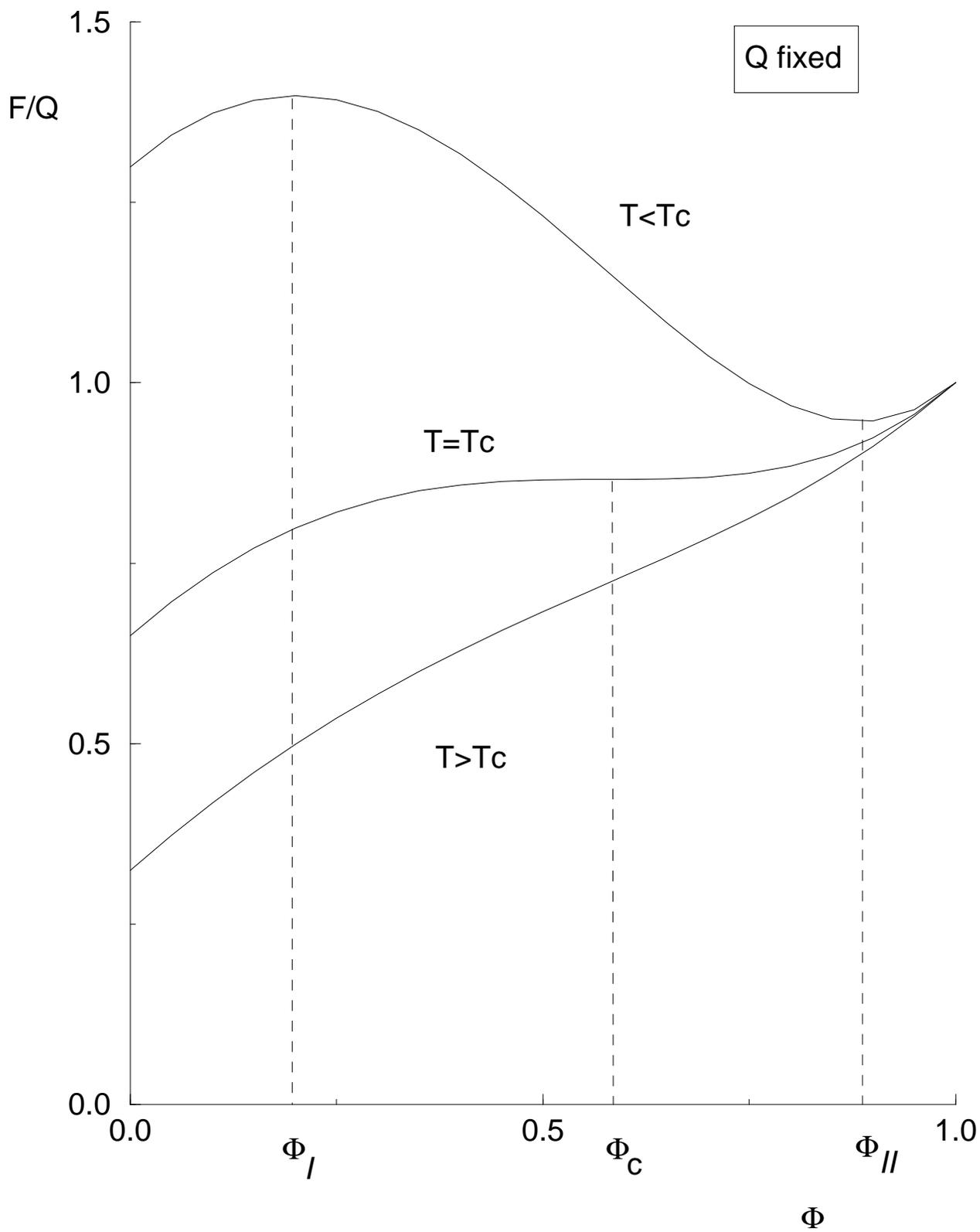
\begin{figure}
\caption{The Helmholtz free energy for a fixed charge $Q>0$. At
temperatures lower that the critical one there is a maximum at
$\Phi_{I}$ and a minimum at $\Phi_{II}$. At $T_c$ both coincide
($\Phi_c=1/\protect\sqrt{3}$ there), and below there is neither
a maximum nor a minimum in the considered range of $\Phi$.}
\label{fig3}
\end{figure}

Further insight can be gained on the stability of the branches $I$ and
$II$ when we consider the Helmholtz
potential (\ref{potencial}) shown in figure (\ref{fig3}). In fact, it
comes out clearly from the figure that the branch $II$ represents a
minimum of the potential, while the branch $I$ is a maximum. Since
black holes in $I$ have a higher mass than those in $II$ they have a
lower ratio $Q/M$ and they are in the branch closer to the uncharged
black holes (i.e. Schwarzschild ones). In fact, this black holes possess
a negative heat capacity that indicate they cannot be held in equilibrium
with an infinite radiation bath. Depending on the sign of the
perturbation they can evolve towards the stable minimum by losing
mass (for instance,
if they keep a temperature higher than that of the bath) or
towards the Schwarzschild--like holes by indefinitely gaining mass
in favor of the radiation bath (for instance,
if they keep a lower temperature than
the bath). On the other hand, black holes in
branch $II$ have a higher ratio $Q/M$ that makes their heat capacity
positive, allowing them to be in equilibrium with radiation. As we rise
the temperature, the two branches merge together and the stable
state $II$ disappears for $T>T_c$, making the hole evolve towards an
infinite mass state.

We can thus compare this phenomenon with what happens in a magnet near
criticality: At high temperature the magnet do not present any net
magnetization $(\vec M)$, but as we lower its temperature below it Curie
point $(T=T_c)$ one observes a spontaneous magnetization to appear even
at zero external applied field. Further lowering its temperature, makes
to increase the magnetization. Two values of the magnetization
$\pm\vec M$ are possible depending on the initial perturbation. The
analogy with our black hole system here is limited since one of the
two branches of the electrostatic field (our order parameter),
the $\Phi_I$, is unstable. In this restricted sense, the transition
between branch $I$ and $II$ ends at $\Phi_c$ which can be called a
critical point. $T_c$ now being the parameter that gives the change in
the stability properties of highly charged--rotating black holes
($\Phi_{II}$).

\section{Thermodynamics of a hole with negative mass?}

In the above analysis we have supposed that $\Phi$ could not be bigger
than one since this represents the extremal black hole. If we also consider
negative charges this statement generalizes to $|\Phi|\leq1$. But let us
be curious enough to see what represents $|\Phi|>1$. To do so, we take
the last equation in (\ref{temperatura}), and plug the explicit form
of $r_+$ into
\begin{equation}
\Phi={Q\over M+\sqrt{M^2-Q^2}}~;
\end{equation}
to keep $\Phi$ real we consider $|M|\geq |Q|$. One can thus probe the
following implications:

\noindent
1) $|\Phi|\leq1\quad\Rightarrow\quad M\geq 0\quad\Rightarrow\quad
{\text{sign}}\{\Phi\}={\text{sign}}\{Q\}$
and

\noindent
2) $|\Phi|\geq1\quad\Rightarrow\quad M\leq0\quad\Rightarrow\quad
{\text{sign}}\{\Phi\}=-{\text{sign}}\{Q\}$

[The analogous implications hold for the rotating (Kerr) hole and can be
probed by use of the second equation in (\ref{temperatura}).]

We are thus dealing with negative mass objects. Let us also note that
the Kerr--Newman metric (\ref{kn}) is invariant under the change of
sign of its parameters and the radial coordinate
\begin{equation}
M_*\to-M~,~~r_*\to-r~,~~Q_*\to-Q~,~~J_*\to-J~.
\label{estrella}\end{equation}

Note that while the last two reflections have a natural interpretation,
we usually do not consider negative mass objects due to its odd stability
properties and the fact that the represent naked singularities
[also, changing the sign of $r$ is not a covariant transformation,
but for our following discussion of the thermodynamical variables this will
not be relevant.]

If we extend the usual thermodynamic relations by the discrete $*$ symmetry
we find that (considering now the quantity $M>0$)
\begin{equation}
r_\pm(M_*)=r_\pm(-M)=-r_\mp(M)~.
\end{equation}
by direct use of the definitions. That is all the dependence on the mass
we need to compute the thermodynamic variables given by
Eqs. (\ref{entropia}) and (\ref{temperatura}).

For the ``intensive'' variables
\begin{equation}T_*=-T(r_-)>T_+>0~;~~
\Omega_*=\Omega(r_-)>\Omega_+>0~;~~
\Phi_*=\Phi(r_-)>\Phi_+>0~,
\label{inten}\end{equation}
and for the ``extensive'' variables
\begin{equation}
0<S_*=S(r_-)<S_+~;~~J_*=-J<0~;~~Q_*=-Q<0~.
\label{exten}\end{equation}
[where the inequalities for $\pm$ quantities
are due to $r_+(M)>r_-(M)$, and they become
equalities in the extremal case.]

We are interested in the case of holes in equilibrium with a radiation
bath, i.e. in the
canonical ensamble. For that, we have to fix the temperature $T$,
the charge $Q$ and the angular momentum $J$. We have thus to impose
only the first two of the $*$ transformations (Eqs. (\ref{estrella}))
to be able to deal with positive and negative mass states on the same
foot at fixed $Q$ and $J$.
As a result, negative mass thermodynamic states will have the
last two quantities in Eqs. (\ref{inten}) and (\ref{exten}) with the
sign opposed to that shown.
Thus, keeping the positiveness of the angular momentum and
charge, but negative angular velocity and electrostatic potential.

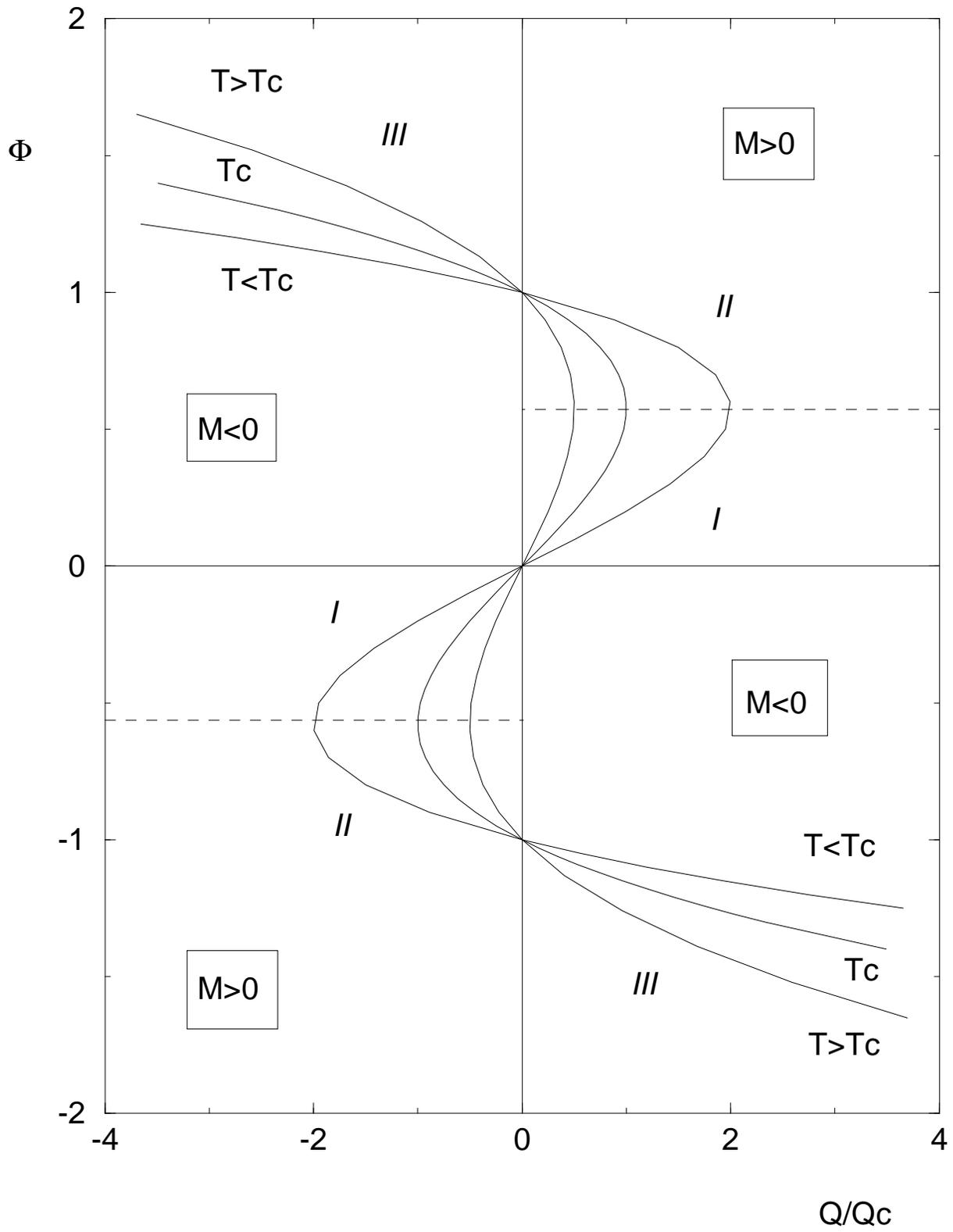
\begin{figure}
\caption{This is the extended equation of state of charged holes
for all values of the electrostatic potential $\Phi$. Here branch
$I$ is unstable, branch $II$ represent metastable states and branch
$III$ (corresponding to negative mass holes) are stable states.
The Maxwell construction would avoid unstable and metastable sectors
by simply joining branches $III$ along the $\Phi$ axis between the
points $\Phi=\pm1$. Otherwise a transition is possible between 
branches $II$ and $III$ up to the critical value
$\Phi_c=\pm1/\protect\sqrt{3}$.}
\label{fig4}
\end{figure}

In the following we will discuss the simpler nonrotating charged case.
Figure (\ref{fig4}) shows how the equation of state extended to all the
four quadrants looks like. The region $0<\Phi<1$ in the first quadrant
is what we have discussed in the last section. Its extention to negative
charges is in the third quadrant. The negative mass states are
characterized by $|\Phi|>1$ and are to the right or to the left
of the abscissa axis
according to its positive or negative charge respectively. We shall
characterize this states by the label $III$. We first observe that at
a given temperature according to the value we fix for the charge of the
system we can have one two or three possible states. At absolute
charges below the
critical one we have three states. In addition to the $I$ and $II$ of
the last section we will also have a negative mass state. For
$|Q|>|Q_c|$ only the state $III$ is possible (while there was not state
in the last section analysis). And finally, for the critical value of the
charge only two states are allowed. The equation of state crosses also
three times the $\Phi$ axis. The points $\Phi=\pm1$ correspond to the
$M\to0$ limit but in such a way that the ratio $Q/M\to\pm1$, thus $Q\to0$
too. The point in the origin corresponds to the infinite mass limit.
Note that the full figure looks like the corresponding one for the
mean field approximation to the Ising model with a non-zero applied
external magnetic field and an infinite critical temperature \cite{P88}.

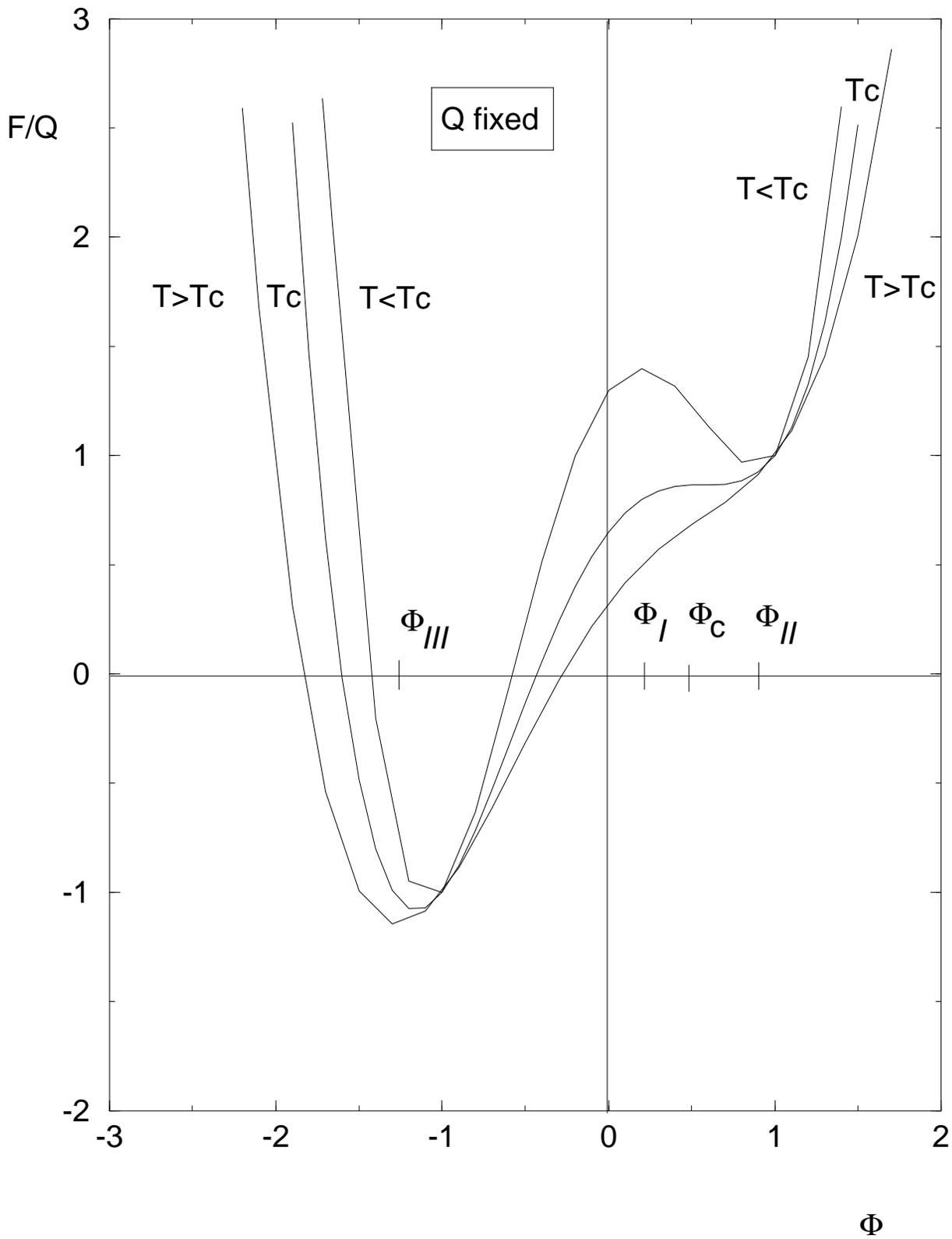
\begin{figure}
\caption{The Helmholtz potential of a Reissner--Nordstr\"om hole
as a function of the electrostatic potential (acting as an order
parameter). At lower temperatures than the critical one there is
a barrier (with a local maximum at $\Phi_I$) separating the
metastable state $\Phi_{II}$ from the ground state (with negative
mass) $\Phi_{III}$. Thus,
a first order transition is needed to pass from one to the
other. As we rise the temperature, that barrier disappears
allowing the transition be of second order. For higher temperatures
only the state $III$ will be stable.}
\label{fig5}
\end{figure}

In figure (\ref{fig5}) we plot the Helmholtz potential that will
help us to discuss the
most relevant question of the stability of the negative mass objects.
For a fixed positive charge $Q$ we have three qualitatively different
kind of potential depending on the value of the temperature. For
temperatures lower than the critical one, the potential has a minimum
in $\Phi_{III}\leq -1$ corresponding to the negative mass sector, then
a relative maximum at $0\leq\Phi_{I}<1/\sqrt{3}$ and a relative minimum
at $1/\sqrt{3}<\Phi_{II}\leq 1$. We thus see that the sector $II$ of
the black holes are now a metastable state that could evolve (through
a first order transition) to a negative mass state. As we rise the
temperature to the critical one this transition will be smooth and
of second order. Above the critical temperature, the metastable state
disappears and only the negative mass object remains, which happens to
be the most stable object of the ensamble. Since this object has a
positive temperature, it could eventually  emit more radiation than
received form the bath, thus decreasing its mass, but as the absolute
value of $M$ increases, this makes the emission temperature to decrease,
below the bath temperature, thus stabilizing the object. Conversely,
a fluctuation in the temperature of the object that make it decrease,
would make it gain positive mass on the bath making its absolute
value decrease and thus finally radiating at a temperature higher than
that of the bath, reaching in that way the stable point again.

\section{Discussion}

In the first part of the paper we have recomputed and corrected the
critical exponents for highly rotating--charged black holes given in
\cite{L95}. In Ref. \cite{L95}
we have used them to guess the effective dimension of the system. This
cannot be done here since we have computed only the thermodynamic
exponents (Eqs. (\ref{st})--(\ref{kx})), which are {\it independent}
of the dimensionality of the system. Critical exponents related to the
two point correlation function {\it do} depend on the dimensionality
and can be used to determine it.

It is worth
noting that critical phenomena have also been observed in numerical
analysis of the collapse of a scalar field to form a black hole
\cite{C93}, and that the classical value 1/2 for the exponents
have also been found there \cite{P95}. 

The exponents we computed are classical. To improve them one should
have a quantum version of the analysis, use the Renormalization
Group and get the corrections to (\ref{st})--(\ref{kx}). This can
be done by considering the effective Hamiltonian derived from the
Helmholtz potential and is under current research. The quantum
analysis could eventually give us a physical interpretation of the
negative energy states.
In conclusion, although we treat semiclassical (neither astrophysical
nor Planck scale) black holes, it seems that some new (perhaps quantum)
phenomena appears below the critical temperature $T_c$.

In Refs. \cite{C79} it was introduced the concept of ``spin entropy''
and ``spin temperature'' for Kerr black holes in terms of the 
quantities defined on the internal horizon and it was studied its
thermodynamics. In particular, $T_-<0$, was interpreted as
reflecting the inversion of population like in usual spin systems.
In our paper, on the other hand, thermodynamic quantities evaluated on
the internal horizon appear as a consequence of extending the usual
formulae for black holes to objects with negative mass described
by the Kerr--Newman geometry. Its temperature is $-T_-$, thus defined
positive, and they can be in equilibrium with an infinite radiation
bath. There still remain, of course, to know if there is any object
in nature that could (at least effectively) be described in such way.

\begin{acknowledgments}
C.O.L was supported by the NSF grant PHY-95-07719 and by research
founds of the University of Utah.

\end{acknowledgments}

\begin{references}

\bibitem{B73}  J. D. Bekenstein,  Phys. Rev., D7, 949 (1973); ibid,
D7, 2333 (1973); ibid, D9, 3292 (1974).\par

\bibitem{H74} S. W. Hawking, Nature., {\bf 248}, 30 (1974);
Commun. Math. Phys., {\bf 43}, 199 (1975).\par

\bibitem{D77} P. C. W. Davies, Proc. R. Soc. Lond., A353, 499 (1977).\par

\bibitem{L93} C. O. Lousto, Nucl. Phys., B410, 155 (1993).
Erratum, ibid, B449, 433 (1995).\par

\bibitem{TL80} D. Tranah and  P. T. Landsberg, Collective Phenomena,
{\bf 3}, 81 (1980).\par

\bibitem{S73} L. Smarr, Phys. Rev. Lett., {\bf 30}, 71 (1973).\par

\bibitem{P88} G. Parisi, {\it Statistical Field Theory} (Addison--Wesley,
Redwood City, CA, 1988), Ch 3.

\bibitem{L95} C.O.Lousto, Phys.Rev., D51, (1995) 1733.\par

\bibitem{C93} M. Choptuik, Phys Rev. Lett, {\bf 70}, 9 (1993).\par

\bibitem{P95} J. Pullin, Phys. Lett, A204, 7 (1995).\par

\bibitem{C79} A. Curir, Nouvo Cim., B51, 262 (1979); Europhys. Lett.,
{\bf 9}, 609 (1989). \par

\end{references}
\end{document}